\def\rfr#1{eq. (\ref{#1})}

\def\vig{``}

\def\bar{\begin{eqnarray}}
\def\ear{\end{eqnarray}}
\def\bb{\bibitem}
\def\eqi{\begin{equation}}
\def\eqf{\end{equation}}
\def\eqia{\begin{eqnarray}}
\def\eqfa{\end{eqnarray}}
\def\rp#1#2{{#1\over#2}}

\def\lb#1{\label{#1}}

\def\oc2{$\mathcal{O}(c^{-2})$}

\def\bds#1{\boldsymbol{#1}}

\documentclass[11pt]{article}
\usepackage{url}
\usepackage{rotating,amsmath,amsthm,amscd,amssymb}
\usepackage{latexsym}
\usepackage{graphicx,epsfig}

\RequirePackage{color}

\begin{document}

\noindent{\bf \LARGE{The effect of general relativity on hyperbolic orbits and its application to the flyby anomaly
}}
\\
\\
\\
{L. Iorio}\\
{\it INFN-Sezione di Pisa. Address for correspondence: Viale Unit$\grave{a}$ di Italia 68\\
70125 Bari, Italy
\\tel./fax 0039 080 5443144
\\e-mail: lorenzo.iorio@libero.it}

\vspace{4mm}
\begin{abstract}
We investigate qualitatively and quantitatively the impact of the general relativistic gravito-electromagnetic forces on hyperbolic orbits around a massive spinning body. The gravito-magnetic field, which is the cause of the well known Lense-Thirring precessions of elliptic orbits, is generated by the spin $\bds S$ of the central body. It deflects and displaces the trajectories differently according to the mutual orientation of $\bds S$ and the orbital angular momentum $\bds L$ of the test particle. The gravito-electric force, which induces the Einstein precession of the perihelion of the orbit of Mercury, always deflects the trajectories inward irrespective of  the $\bds L-\bds S$ orientation. We numerically compute their effect on the range $r$, radial and transverse components $v_r$ and $v_{\tau}$ of the velocity and speed $v$ of the NEAR spacecraft at its closest approach with the Earth in January 1998  when it experienced an anomalous increase of its asymptotic outgoing velocity $v_{\infty\ {\rm o}}$ of  $13.46\pm 0.01$ mm sec$^{-1}$; while the gravito-electric force was modeled in the software used to process the NEAR data, this was not done for the gravito-magnetic one.  The range-rate and the speed are affected by general relativistic gravito-electromagnetism at $10^{-2}\ (\rm gravito-electric)-10^{-5}\ (\rm gravito-magnetic)$  mm sec$^{-1}$ level. The changes in the range are of the order of $10^{-2}\ (\rm gravito-magnetic)-10^1\ (\rm gravito-electric)$ mm.

\end{abstract}

 \section{Introduction}
 In this paper we will investigate the effects of general relativity, in its weak-field and slow-motion approximation,
 on unbound, hyperbolic orbits of test particles approaching a body of mass $M$ and angular momentum $\bds S$.
  We will consider  both the gravito-magnetic  and the  gravito-electric relativistic forces induced by the stationary and static components, respectively, of the field of $M$ \cite{MashNOVA}.

The results obtained will be applied to some realistic planet-spacecraft scenarios in the Solar System to see if the predicted effects fall within the current or future sensitivity level of the ranging techniques.
In particular, we will examine the so-called flyby anomaly consisting of  unexplained changes of the asymptotic outgoing velocities of some spacecraft (Galileo, NEAR, Cassini and MESSENGER) that occurred at their closest approaches with the Earth \cite{And07,And08}.

\begin{figure} \includegraphics[width=\columnwidth]{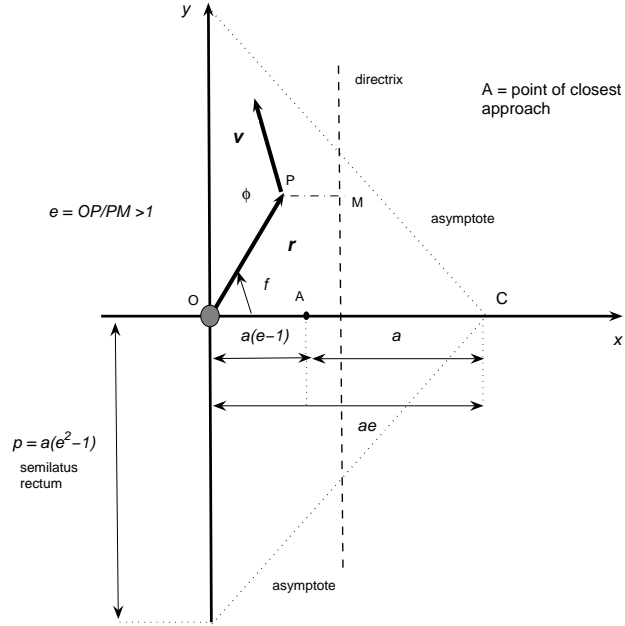}
 \caption{\label{hyperbola} Essential features of the unperturbed Newtonian hyperbola. P is the position of a test particle at time $t$ with respect to the body located at the origin of the chosen reference frame. The smallest distance is $r_{\rm min}=a(e-1)$.
 The angle between $\bds r$ and $\bds v$ is $\phi$. M here is not to be confused with the central body of mass $M$.}
\end{figure}

Let us recall some basics of the Newtonian hyperbolic orbit \cite{Roy} which represents, in this case, the reference unperturbed path; see Figure \ref{hyperbola}.
%
%
Its equation is
\eqi r = \rp{p}{1+e\cos f},\eqf
where $p=a(e^2-1)$ is the semilatus rectum, and  $f$ is the true anomaly considered  positive in the anti-clockwise direction from the point of closest approach A;  $a$ is the semi-major axis and $e>1$.
The speed is
\eqi v =\sqrt{GM\left(\rp{2}{r} + \rp{1}{a}\right)},\eqf so that
\eqi v_{\infty}=\sqrt{\rp{GM}{a}}.\eqf
The asymptotic ingoing and outgoing speeds, which are equal to $v_{\infty}$ in the unperturbed case, are denoted with $v_{\infty\ {\rm i}}$ and $v_{\infty\ {\rm o}}$, respectively.
The angle $\phi$ between the $\bds r$ and $\bds v$  is defined by
\eqi\sin\phi = \rp{1+e\cos f}{\sqrt{1+e^2+2e\cos f}},\eqf
\eqi\cos\phi = -\rp{e\sin f}{\sqrt{1+e^2+2e\cos f}}.\eqf
For $x=0$, $y=\pm p$, i.e. $f=\overline{f}=\pm \pi/2$,
it is
\eqi v_{x0} = \overline{v}\sin\overline{\phi}=\sqrt{\rp{GM}{p}},\eqf
\eqi v_{y0} = \overline{v}\cos\overline{\phi}=\mp e\sqrt{\rp{GM}{p}}.\eqf
 \section{The gravito-magnetic  force}
 In this section we will deal with the effect of the  general relativistic gravito-magnetic  force on the hyperbolic motion of a test particle approaching a spinning body of mass $M$ and angular momentum $\bds S$.

 Let us briefly recall that, in the weak-field and slow-motion linear approximation of general relativity, the off-diagonal components $g_{0i}, i=1,2,3$ of the space-time metric tensor, related to the mass-energy currents of the source,  induce  a gravito-magnetic field $\bds B_{\rm g}$ \cite{MashNOVA} by analogy with the magnetic field caused by moving electric charges in the linear Maxwellian electromagnetism. Far from an isolated rotating body, the gravito-magnetic field becomes  \cite{Lic}
 \eqi \bds B_{\rm g} = -\rp{G}{c r^3}\left[\bds S -3\left(\bds S\bds\cdot\bds{\hat{r}}\right)\bds{\hat{r}}\right],\lb{BGM}\eqf   where $G$ is the Newtonian gravitational constant and $c$ is the speed of light in vacuum. It exerts   the non-central Lorentz-like acceleration  \cite{MashNOVA}
 \eqi \bds A^{\rm GM} = -\rp{2}{c}{\bds v}\bds\times\bds B_{\rm g}\lb{acc_gm}\eqf
 upon a test particle moving with velocity $\bds v$.
 For ordinary astronomical bodies like, e.g., the Earth and the Sun, $A^{\rm GM}$  is many orders of magnitude smaller than the Newtonian monopole $A^{\rm N} = {GM}/{r^2}$, so that it can be considered as a small perturbation.

 The action of \rfr{acc_gm} in the case of unperturbed close orbits, giving rise to, e.g., the Lense-Thirring precession  of the ellipse of a test particle \cite{LT,Mash84,Ruf03}, has been the subject of intense activity, both theoretically and observationally, in recent times \cite{IorNOVA}.

 Here we will consider as reference path of a Newtonian hyperbolic trajectory. In order to work out the effects of the gravito-magnetic field on it, we will numerically integrate the equations of motion in cartesian rectangular coordinates \cite{Sof89} for some particular orbital geometries over a time span including the epoch of closest approach to $M$ which is assumed located at the origin of the coordinate system.

\subsection{Qualitative features for equatorial and polar osculating orbits}
\begin{figure}
\includegraphics[width={3cm}]{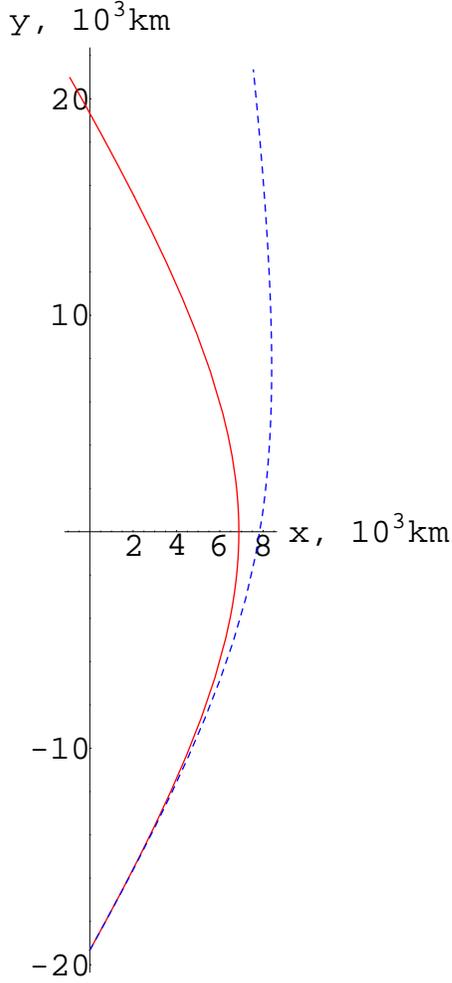}
 \caption{\label{LT1} Effect of the gravito-magnetic force on the hyperbolic motion of a test particle around an astronomical rotating body located at the origin of the depicted frame. The body's spin $\bds S$ is directed along the positive $z$ axis, i.e outside the figure. Red line: unperturbed hyperbola. Blue dashed line: perturbed orbit. For illustrative purposes we choose the Earth as central body and re-scaled the magnitude of its gravito-magnetic force by $10^{10}$ so that $A^{\rm GM}/A^{\rm N}=0.4$ at perigee.  We adopted the initial conditions $x_0=0,\ y_0=-p=-a(e^2-1),\ z_0 = 0,\ v_{x0}>0,\ v_{y0}>0,\ v_{z0} = 0$; the particle moves  in the equatorial plane of the spinning Earth in such a way that the radius vector rotates in the same sense with respect to the Earth, i.e anticlockwise. We used $a=8493.326$ km, $e=1.81$. The perturbed orbit is deflected outward with respect to the unperturbed one.}
\end{figure}

\begin{figure}
\includegraphics[width={10cm}]{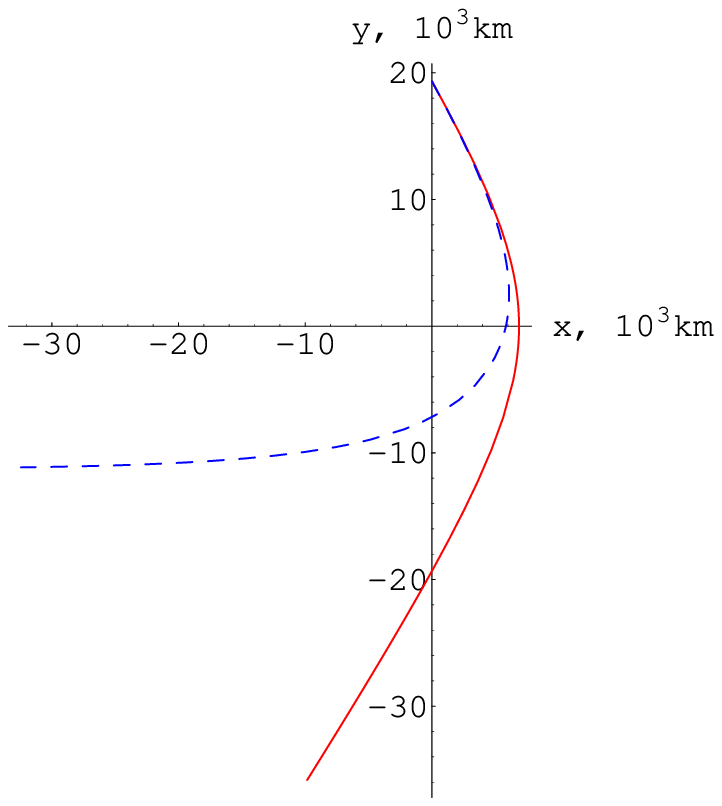}
 \caption{\label{LT2} Effect of the gravito-magnetic force on the hyperbolic motion of a test particle around an astronomical rotating body located at the origin of the depicted frame. The body's  spin $\bds S$ is directed along the positive $z$ axis, i.e outside the figure. Red line: unperturbed hyperbola. Blue dashed line: perturbed orbit. For illustrative purposes we choose the Earth as central body and re-scaled the magnitude of its gravito-magnetic force by $10^{10}$ so that $A^{\rm GM}/A^{\rm N}=0.4$ at perigee.  We adopted the initial conditions $x_0=0,\ y_0=p=a(e^2-1),\ z_0 = 0,\ v_{x0}>0,\ v_{y0}<0,\ v_{z0} = 0$; the particle moves  in the equatorial plane of the spinning Earth in such a way that the radius vector rotates in the opposite sense with respect to the Earth, i.e. clockwise. We used $a=8493.326$ km, $e=1.81$. The perturbed orbit is deflected inward with respect to the unperturbed one.}
\end{figure}

First, we will consider a trajectory lying in the equatorial plane of the rotating body for the cases of co-rotation (Figure \ref{LT1}) and counter-rotation (Figure \ref{LT2}) of the particle's radius vector $\bds r$ with respect to the diurnal rotation of $M$ whose spin $\bds S$ is assumed to be directed along the positive $z$ axis (anti-clockwise diurnal rotation).
%
%
%
%
%
%
Just for illustrative purposes, we assumed the Earth as source of the gravitational field and re-scaled its gravito-magnetic force by a factor $10^{10}$ in such a way to still keeping the condition $A^{\rm GM}/A^{\rm N}<1$ valid over the entire orbit.
It turns out that the perturbed trajectory remains confined in the equatorial plane of the central body;
 for co-rotation the path is deflected outward with respect to the unpertubed hyperbola, while for counter-rotation it is deflected inward. Indeed, for equatorial orbits $\bds A^{\rm GM}$ is entirely in-plane because $\bds B_{\rm g}$ is directed along the negative $z$ axis; for co-rotating particles it is radially directed outward at the point of closest approach and decreases the gravitational pull felt by the orbiter, while for counter-rotating probes it is radially directed inward at the pericentre and increases the net gravitational acceleration. The flyby epoch is left almost unaffected. By taking the difference between the integrated perturbed and unperturbed orbits sharing the same initial conditions, it can be shown that in the co-rotating case the radial velocity, which is one of the direct observables in real planet-spacecraft close encounters, experiences an increase with respect to the unperturbed one just around the flyby epoch, while the radial components of $v_{\infty\ {\rm i}}$ and  $v_{\infty\ {\rm o}}$ are left unaffected. In the counter-rotating case $v_r$ decreases at the closest approach. Concerning the body-centric range $r$, it turns out that it suddenly increases (decreases) around the flyby epoch for the co-(counter-)rotating case and remains about at that level also after the flyby.

\begin{figure}
\includegraphics[width={10cm}]{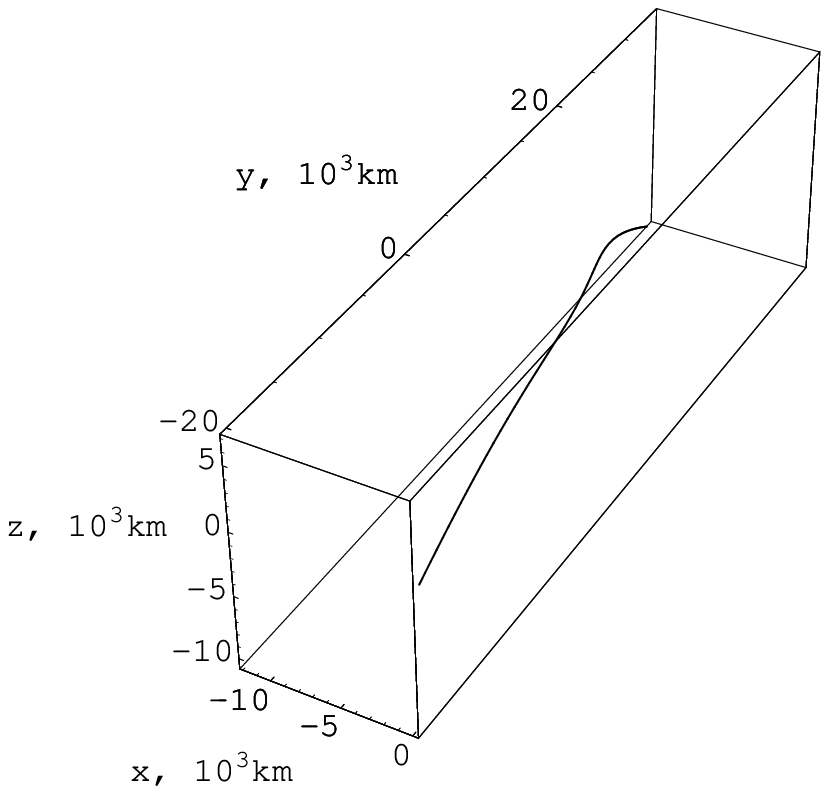}
 \caption{\label{LT3} Effect of the gravito-magnetic force on the hyperbolic motion of a test particle around an astronomical rotating body located at the origin of the depicted frame. The body's spin $\bds S$ is directed along the positive $z$ axis, i.e outside the figure.  For illustrative purposes we choose the Earth as central body and re-scaled the magnitude of its gravito-magnetic force by $10^{10}$ so that $A^{\rm GM}/A^{\rm N}=0.4$ at perigee.  We adopted the initial conditions $x_0=0,\ y_0=-p=-a(e^2-1),\ z_0 = 0,\ v_{x0}=0,\ v_{y0}>0,\ v_{z0} > 0$ to have the spacecraft initially moving in the osculating $\{yz\}$ plane. The perturbed trajectory is displaced along the $x$ axis on the upper left corner. }
\end{figure}

\begin{figure}
\includegraphics[width={10cm}]{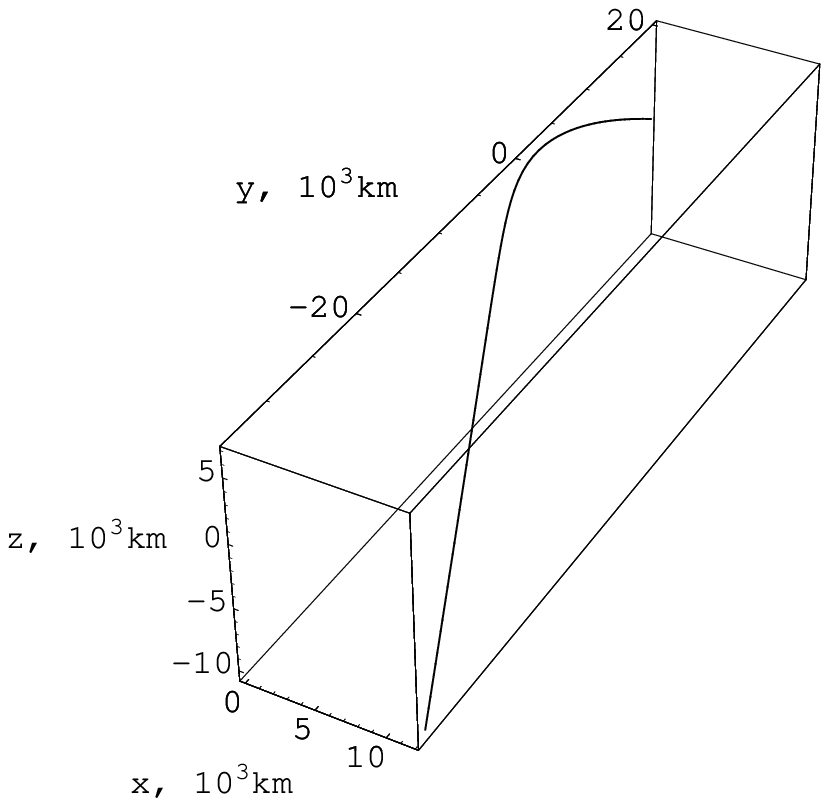}
 \caption{\label{LT4} Effect of the gravito-magnetic force on the hyperbolic motion of a test particle around an astronomical rotating body located at the origin of the frame shown with its spin directed along the positive $z$ axis, i.e outside the figure.  For illustrative purposes we choose the Earth as central body and re-scaled the magnitude of its gravito-magnetic force by $10^{10}$ so that $A^{\rm GM}/A^{\rm N}=0.4$ at perigee.  We adopted the initial conditions $x_0=0,\ y_0=p=a(e^2-1),\ z_0 = 0,\ v_{x0}=0,\ v_{y0}<0,\ v_{z0} > 0$ to have the spacecraft initially moving in the osculating $\{yz\}$ plane. The perturbed trajectory is displaced along the $x$ axis on the lower right corner. }
\end{figure}

Let us, now, consider the case in which the unperturbed hyperbola entirely lies in an azimuthal plane, e.g. the $\{yz\}$ plane, containing the spin $\bds S$ of the central body. Now, since $\bds v$ is contained in the same plane of $\bds B_{\rm g}$, the gravito-magnetic acceleration is out-of-plane, so that it can be expected that the perturbed trajectory will be displaced along the $x$ axis. This fact is confirmed by a numerical integrations shown in Figure \ref{LT3}  and Figure \ref{LT4}
from which the displacement of the orbit perpendicularly to the initial osculating plane is apparent.
\subsection{The flyby anomaly: the NEAR case}
\begin{figure}
\includegraphics[width={7cm}]{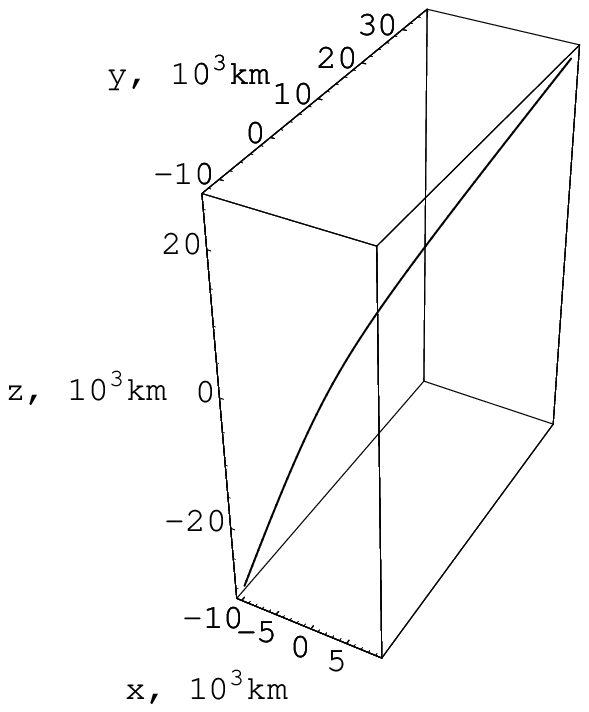}
 \caption{\label{NEAR} Unperturbed hyperbola of NEAR; its osculating plane is tilted by 108 deg to the $\{x,y\}$ plane assumed coincident with the Earth's equator. The starting point is in the right upper corner ($x_0>0,\ y_0>0,\ z_0>0$). The Earth has to be imagined located at the origin of the coordinate system with its spin $\bds S$ directed along the positive $z$ axis. }
\end{figure}

Many spacecraft launched in interplanetary missions make use of one or more Earth's flybys in order to gain or lose the heliocentric energy required to reach their far targets (planets, asteroids, comets) without using huge and expensive amounts of propellant \cite{Flan,van}.  In the case of GALILEO (twice), NEAR, Cassini and MESSENGER  unexplained variations in $v_{\infty}$ were detected \cite{And07,And08}; the largest one was measured at the close encounter of  NEAR  with the Earth that occurred in 1998 and amounts to \eqi \Delta v_{\infty} = 13.46\pm 0.01\ {\rm mm\ sec}^{-1}.\eqf
The unperturbed hyperbola of NEAR is depicted in Figure \ref{NEAR}.

Anderson et al. in \cite{And08} derived an empirical formula which satisfactorily fit all the six flyby anomalies measured so far.
It is
\eqi \rp{\Delta v_{\infty}}{v_{\infty}} = \left(\rp{2\omega R}{c}\right)\left(\cos\delta_{\rm i}-\cos\delta_{\rm o}\right)=\left(3.099\times 10^{-6}\right)\left(\cos\delta_{\rm i}-\cos\delta_{\rm o}\right),\lb{andi}\eqf
where $\omega$ and $R$ are the Earth's angular speed and equatorial radius, respectively, while $\delta_{\rm i}$ and $\delta_{\rm o}$ are the ingoing and ongoing geocentric declinations, respectively.
Concerning possible explanations in terms of known physics, many dynamical effects (tides, atmospheric drag, Earth's albedo, solar wind, terrestrial magnetic field, spin-rotation coupling between electromagnetic waves and spacecraft and Earth rotations) were preliminarily considered by L\"{a}mmerzahl et al. in \cite{Lamm}; an order-of-magnitude approach was followed by confronting the magnitude of the accelerations induced by  standard forces considered with the one which is assumed to be responsible of the flyby anomaly, i.e. $\approx 10^{-4}$ m sec$^{-2}$. As a consequence, all the effects investigated were discarded. However, we note that detailed analyses are in order: indeed, even if some dynamical effect, standard or not, was found to produce an acceleration with the right order of magnitude, it might happen that its signature on the observable quantities  is not correct; that is, it may induce, for instance, a decrease of the radial velocity.
Mbelek in \cite{Mbe08}  suggested that the special relativistic Doppler effect may explain the formula by Anderson et al. \cite{And08}.
Among various explanations in terms of non-conventional physics put forth so far, McCulloch in \cite{McC08} proposed a mechanism based on the hypothesis that inertia is due to a form of Unruh radiation and varies with acceleration due to a Hubble-scale Casimir effect. It qualitatively reproduces the latitude-dependence of \rfr{andi} and is quantitatively in agreement with three of the six measured flybys.

Since \rfr{andi} contains a term including quantities like the speed of light and the first power of the Earth's angular speed which enters general relativistic gravito-magnetic effects, it seems interesting to  apply our previous results concerning the influence of $\bds B_{\rm g}$ on hyperbolic orbits to the NEAR's flyby. Note that the gravito-magnetic force was not modeled in the software used to process the NEAR data.
By using the state vector of NEAR at the flyby epoch (J D Anderson, private communication to the author, November 2008) referred to a geocentric equatorial frame $\{x,y,z\}$,
and \rfr{acc_gm} it turns out that  at the point of closest
approach to Earth along the flyby trajectory
\eqi A_x^{\rm GM} = 3.3\times 10^{-10}\ {\rm m\ sec}^{-2},\eqf
\eqi A_y^{\rm GM} = 7.5\times 10^{-11}\ {\rm m\ sec}^{-2},\eqf
\eqi A_z^{\rm GM} = -1.7\times 10^{-10}\ {\rm m\ sec}^{-2},\eqf
so that
\eqi  A^{\rm GM} = 3.8\times 10^{-10}\ {\rm m\ sec}^{-2}.\eqf

We will now use a numerical integration of the equations of motion perturbed by \rfr{acc_gm}.
We look at a time span starting from the flyby epoch and extending in the future after it: the chosen initial conditions are in Table \ref{statevec}.
\begin{sidewaystable}
\caption{Initial conditions used for NEAR obtained with the HORIZONS software by NASA, JPL at 1998-Jan-23 07:00:00 CT (Coordinate Time, defined
as the uniform time scale and independent variable of the ephemerides http://ssd.jpl.nasa.gov/?horizons$\_$doc$\#$timesys), i.e. 1353 sec before the flyby. Reference frame: ICRF/J2000.0. Coordinate system: Earth Mean Equator and Equinox of Reference Epoch.\label{statevec}
}
\centering
\bigskip
\begin{tabular}{cccccc}
\hline\noalign{\smallskip}
$x_0$ (km) & $y_0$ (km) & $z_0$ (km) & $v_{x0}$ (km sec$^{-1}$) & $v_{y0}$ (km sec$^{-1}$) &  $v_{z0}$ (km sec$^{-1}$) \\
\noalign{\smallskip}\hline\noalign{\smallskip}
 4,496.885594909381  &  6,930.477153733549 & 13,199.11503591246 &  -1.712684317202157  & -8.679677119077454  &  -4.455285829060190 \\
\noalign{\smallskip}\hline
\end{tabular}
\end{sidewaystable}
We will consider the changes of the velocity along the radial $\bds{\hat{r}}$, transverse $\bds{\hat{\tau}}$ and out-of-plane $\bds{\hat{\nu}}$ directions; $\bds{\hat{\nu}}$ is directed along the orbital angular momentum, while $\bds{\hat{\tau}}=\bds{\hat{\nu}}\bds\times\bds{\hat{r}}$ is not directed, in general, along $\bds v$.

\begin{figure}
\includegraphics[width=\columnwidth]{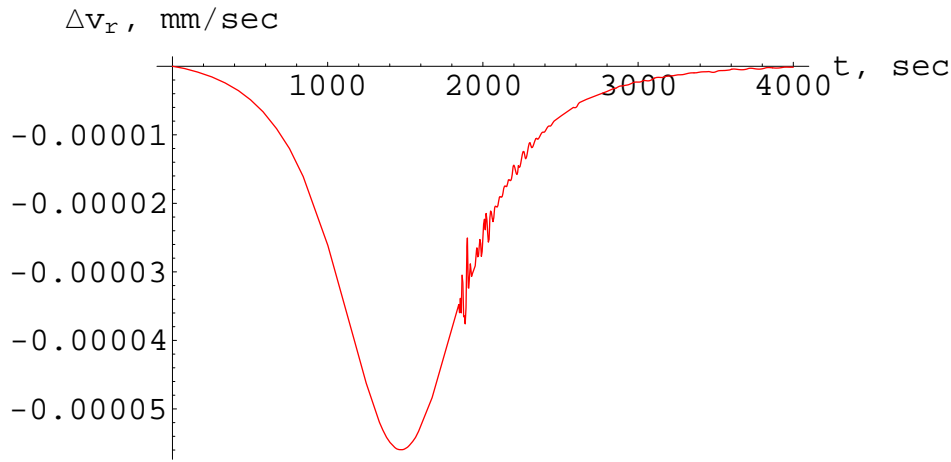}
 \caption{\label{Dvr_LT} Change in the radial velocity $\Delta v_r$ induced by the gravito-magnetic force on NEAR at the Earth's flyby (1998-Jan-23  07:22:56 Coordinate Time (CT)). It has been obtained by taking the difference between the integrated perturbed and unperturbed trajectories sharing the same initial conditions of Table \ref{statevec}. They have been obtained with the HORIZONS software by NASA JPL at 1998-Jan-23 07:00:00 CT and correspond to an instant 1353 sec before the flyby. Reference frame: ICRF/J2000.0. Coordinate system: Earth Mean Equator and Equinox of Reference Epoch. The maximum effect $\Delta v_r^{\rm max} = -5\times 10^{-5}$ mm sec$^{-1}$ occurred just at the flyby. }
\end{figure}

\begin{figure}
\includegraphics[width=\columnwidth]{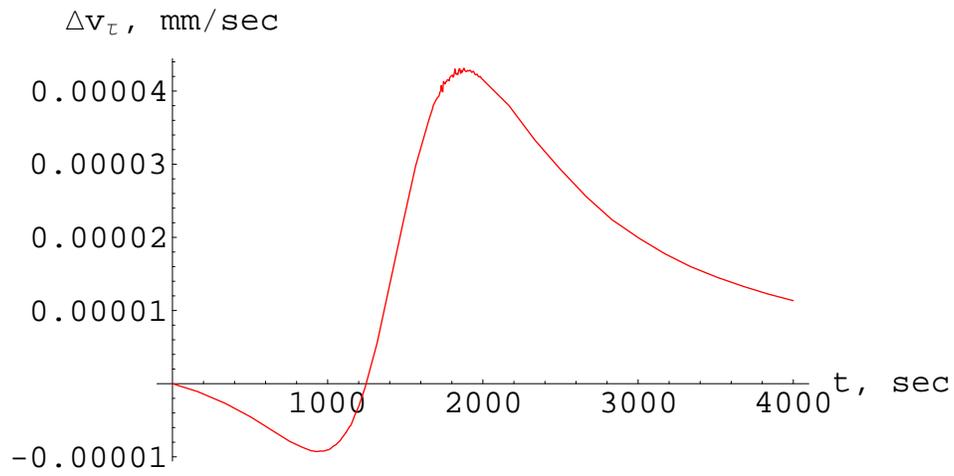}
 \caption{\label{Dvt_LT} Change in the transverse velocity $\Delta v_{\tau}$ induced by the gravito-magnetic force on NEAR at the Earth's flyby (1998-Jan-23  07:22:56 CT). 
 The maximum value reached is of the order of $10^{-5}$ mm sec$^{-1}$. }
\end{figure}

\begin{figure}
\includegraphics[width=\columnwidth]{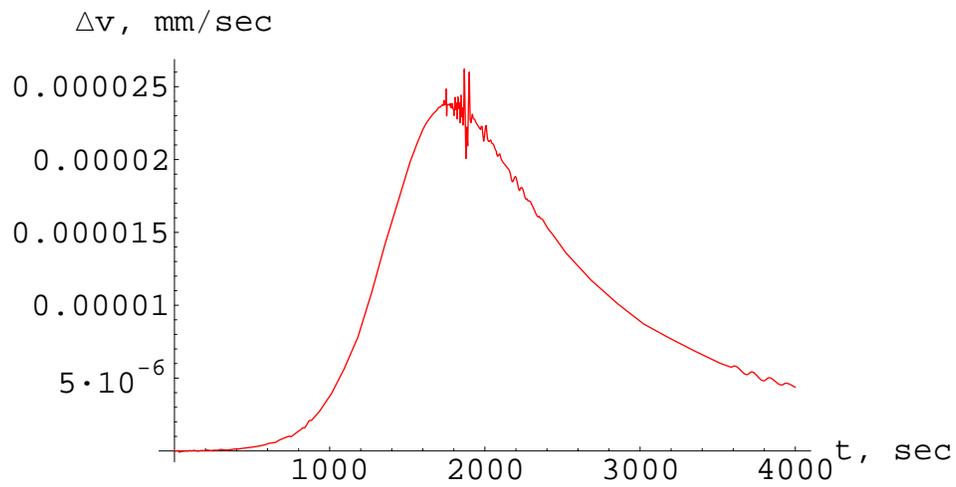}
 \caption{\label{Dv_LT} Change in the  speed $\Delta v$ induced by the gravito-magnetic force on NEAR at the Earth's flyby (1998-Jan-23  07:22:56 CT). 
 The maximum effect $\Delta v^{\rm max} = 2\times 10^{-5}$ mm sec$^{-1}$ occurred just at the flyby while $v_{\infty\ {\rm o}}$ is left almost unchanged. }
\end{figure}

\begin{figure}
\includegraphics[width=\columnwidth]{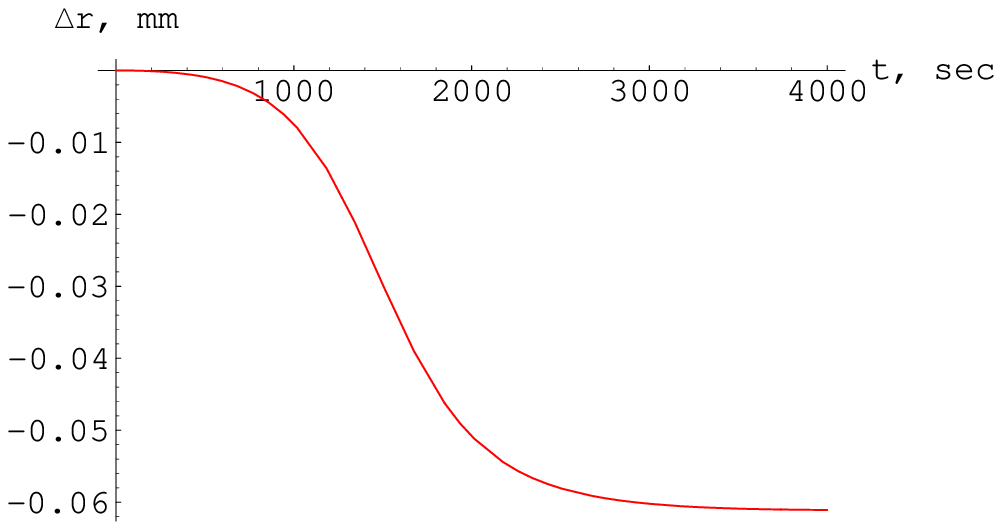}
 \caption{\label{Dr_LT} Change in the radial distance $\Delta r$ induced by the gravito-magnetic force on NEAR at the Earth's flyby (1998-Jan-23  07:22:56 CT). 
 The maximum effect $\Delta r^{\rm max} = -6\times 10^{-2}$ mm took place after the flyby.   }
\end{figure}

The results for $\Delta v_r$, $\Delta v_{\tau}$, $\Delta v$ and $\Delta r$ are shown in Figure \ref{Dvr_LT}-Figure \ref{Dr_LT}, respectively.
They have been obtained by subtracting the unperturbed orbit from the perturbed one, both numerically integrated with the initial conditions
of Table \ref{statevec}.
The gravito-magnetic force of the Earth decreased the radial velocity of NEAR by $\approx 10^{-5}$ mm sec$^{-1}$ just at the flyby, while the transverse velocity was augmented after the flyby up to $10^{-5}$ mm sec$^{-1}$ level; it turns out that the normal velocity was affected at an even smaller level. The total speed $v$ was increased up to $10^{-5}$ mm sec$^{-1}$ i.e. six orders of magnitude smaller than the observed increment. The geocentric range of the spacecraft was reduced by about $10^{-2}$ mm.

\section{The gravito-electric force}
\begin{figure}
\includegraphics[width={3cm}]{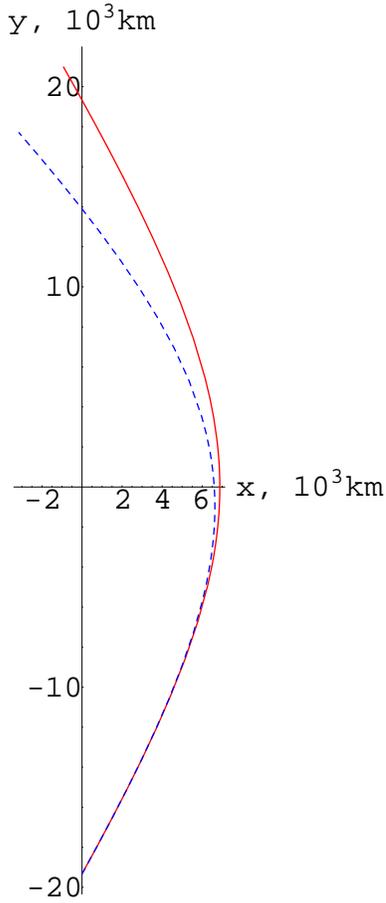}
 \caption{\label{GE1} Effect of the gravito-electric force on the hyperbolic motion of a test particle around an astronomical body located at the origin of the frame shown. Red line: unperturbed hyperbola. Blue dashed line: perturbed orbit. For illustrative purposes we choose the Earth as central body and re-scaled the magnitude of its gravito-electric force by $10^{8}$ so that $A^{\rm GE}/A^{\rm N}=0.1$ at perigee.  We adopted the initial conditions $x_0=0,\ y_0=-p=-a(e^2-1),\ z_0 = 0,\ v_{x0}>0,\ v_{y0}>0,\ v_{z0} = 0$.  We used $a=8493.326$ km, $e=1.81$. The perturbed orbit is deflected inward with respect to the unperturbed one.}
\end{figure}

In the Post-Newtonian approximation of order $\mathcal{O}(c^{-2})$, the acceleration induced by the gravito-electric component of the field of a static mass $M$ is, in standard isotropic coordinates, \cite{Sof89}
\eqi\bds A^{\rm GE} = \rp{GM}{c^2}\left[\rp{4GM}{r^4}\bds r - \rp{v^2}{r^3}\bds r + \rp{4(\bds r\bds\cdot\bds v)}{r^3}\bds v\right].\lb{accge}\eqf
Concerning the influence of \rfr{accge} on the hyperbolic motion, it has only radial and transverse components, so that no departures from the osculating plane occurs. From Figure \ref{GE1} it turns out that the test particle is deflected inward with respect to the unperturbed hyperbola.

The gravito-electric acceleration experienced by NEAR   at the point of closest
approach to earth along the flyby trajectory was
\eqi A_x^{\rm GE} = 9.5\times 10^{-10}\ {\rm m\ sec}^{-2},\eqf
\eqi A_y^{\rm GE} = -5.26\times 10^{-9}\ {\rm m\ sec}^{-2},\eqf
\eqi A_z^{\rm GE} = 3.42\times 10^{-9}\ {\rm m\ sec}^{-2},\eqf
so that
\eqi  A^{\rm GE} = 6.35\times 10^{-9}\ {\rm m\ sec}^{-2}.\eqf

\begin{figure}
\includegraphics[width=\columnwidth]{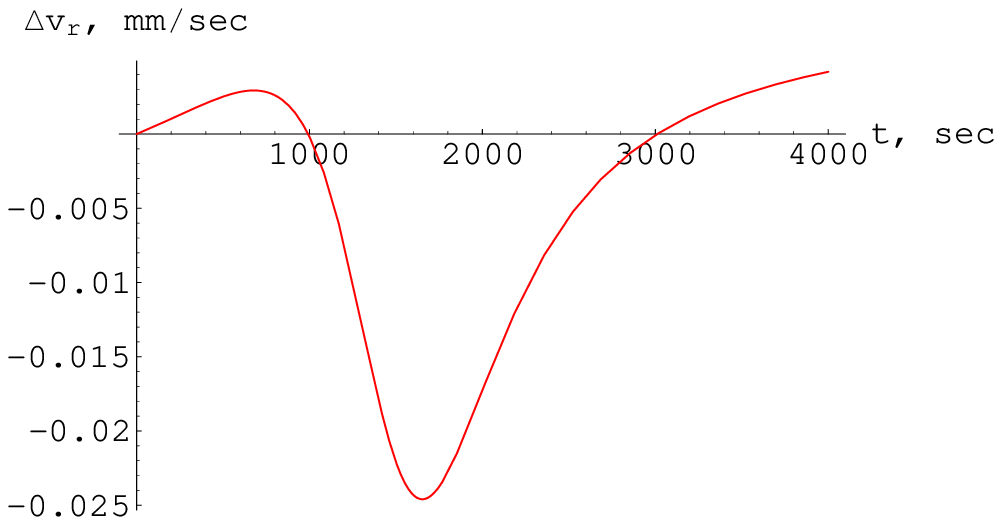}
 \caption{\label{Dvr_SCHW} Change in the radial velocity $\Delta v_r$ induced by the gravito-electric force on NEAR at the Earth's flyby (1998-Jan-23  07:22:56 CT). It is the difference between the integrated perturbed and unperturbed trajectories sharing the same initial conditions of Table \ref{statevec}. They have been obtained with the HORIZONS software by NASA JPL at 1998-Jan-23 07:00:00 CT and correspond to an instant 1353 sec before the flyby. Reference frame: ICRF/J2000.0. Coordinate system: Earth Mean Equator and Equinox of Reference Epoch. }
\end{figure}

\begin{figure}
\includegraphics[width=\columnwidth]{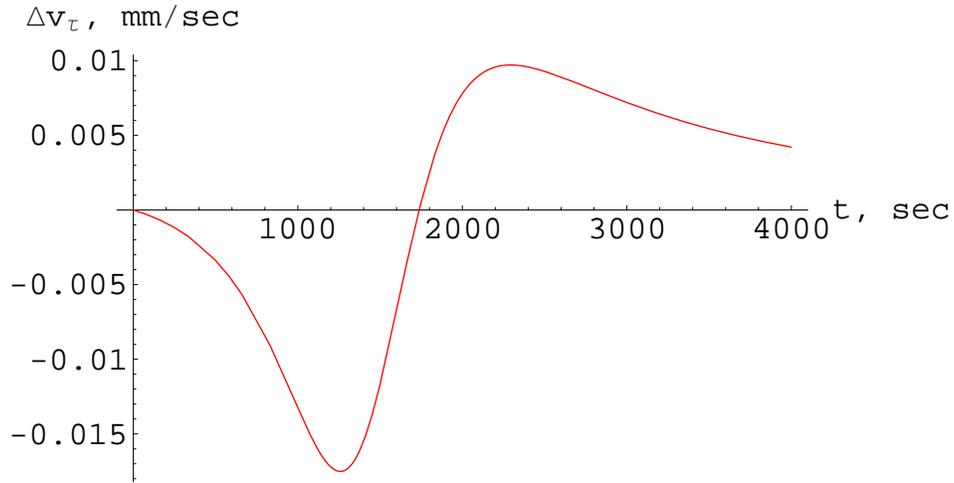}
 \caption{\label{Dvt_SCHW} Change in the transverse velocity $\Delta v_{\tau}$ induced by the gravito-electric force on NEAR at the Earth's flyby (1998-Jan-23  07:22:56 CT). 
 }
\end{figure}

\begin{figure}
\includegraphics[width=\columnwidth]{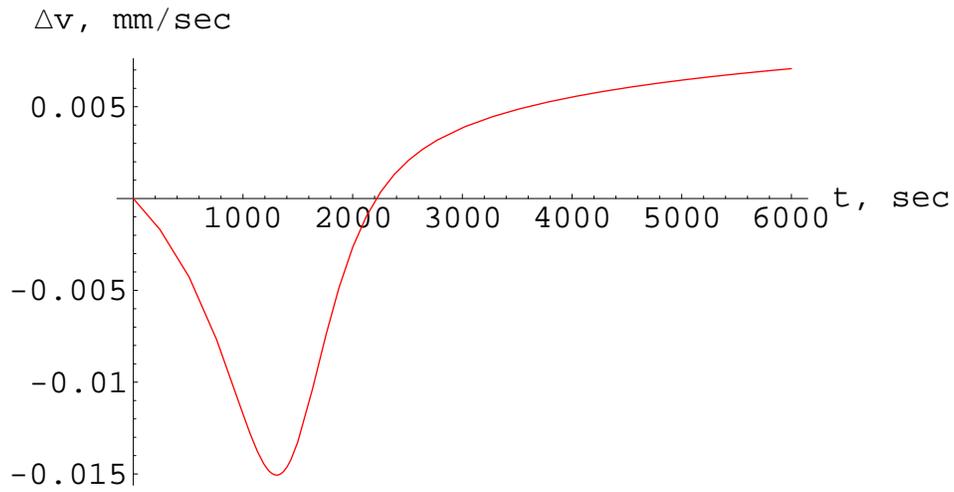}
 \caption{\label{Dv_SCHW} Change in the speed $\Delta v$ induced by the gravito-electric force on NEAR at the Earth's flyby (1998-Jan-23  07:22:56 CT). 
 }
\end{figure}

\begin{figure}
\includegraphics[width=\columnwidth]{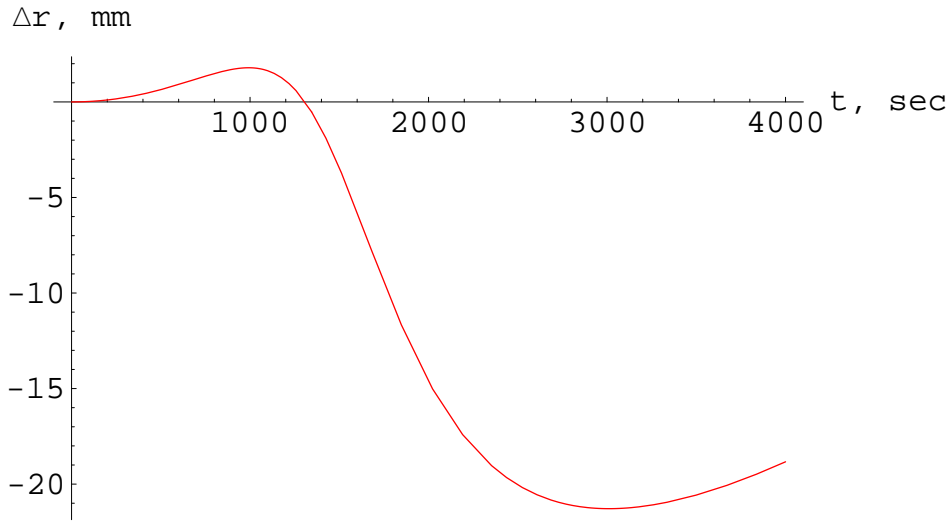}
 \caption{\label{Dr_SCHW} Change in the radial distance $\Delta r$ induced by the gravito-electric force on NEAR at the Earth's flyby (1998-Jan-23  07:22:56 CT). 
 }
\end{figure}

The impact of the gravito-electric force\footnote{Contrary to the gravito-magnetic one, it was modeled in the software used for processing the data.} on $v_r$, $v_{\tau}$, $v$ and $r$ of NEAR at its flyby are depicted in Figure \ref{Dvr_SCHW}-Figure \ref{Dr_SCHW}, respectively; their patterns are quite different from the gravito-magnetic ones; the outgoing asymptotic velocity is changed by an amount of the order of $10^{-3}$ mm sec$^{-1}$. The maximum variations of the range rate, the transverse velocity, the speed and the range are of the order of $10^{-2}$ mm sec$^{-1}$ and $10^1$ mm, respectively.
 \section{Discussions and conclusions}
We investigated qualitatively and quantitatively the impact of general relativity, in its weak-field and slow-motion approximation, on unbound hyperbolic orbits around a massive spinning body.
We considered both the gravito-magnetic and the gravito-electric terms; the first one is responsible of the Lense-Thirring precessions of elliptic orbits, while the second one causes the well known Einstein precession of the perihelion of Mercury of 43.98 arcsec cty$^{-1}$.
The  gravito-magnetic force deflects an equatorial trajectory inward or outward with respect to the unperturbed hyperbola according to the mutual orientation of the orbital angular momentum $\bds L$ of the particle with respect to  the spin $\bds S$ of the central body. For osculating orbits lying in a plane which contains $\bds S$ there is also a displacement in the out-of-plane direction. The gravito-electric force is not sensitive to the $\bds L-\bds S$ orientation and deflects the trajectory inward with respect to the unperturbed hyperbola.

We applied our results to the flyby anomaly experienced by the NEAR spacecraft
at its close encounter with the Earth on January 1998 when its asymptotic outgoing velocity $v_{\infty\ {\rm o}}$ was found larger than the ingoing one by $13.46\pm 0.01$ mm sec$^{-1}$; contrary to the gravito-electric force, the gravito-magnetic one was not modeled in the software used to process the NEAR data. From numerical integrations of the perturbed equations of motion in a geocentric equatorial frame with rectangular cartesian coordinates over a time span extending in the future after the flyby epoch,  we quantitatively investigated the changes in the radial and transverse components of the velocity $v_r$ and $v_{\tau}$, the speed $v$ and the range $r$ of NEAR induced by the general relativistic gravito-electromagnetic forces. Concerning the range, its variations are at the $10^{-2}$ level for the gravito-magnetic force and $10^1$ mm level for the gravito-electric one. The radial and transverse velocities and the speed are affected at  $10^{-2}\ (\rm gravito-electric)-10^{-5}\ (\rm gravito-magnetic)$ mm sec$^{-1}$ level.


\end{document}